\documentclass[11pt,oneside]{article}
\pdfoutput=1

\usepackage{graphicx}

\usepackage{geometry}
\usepackage[font={small,stretch=1.1},labelfont=bf]{caption}

\usepackage[dvipsnames]{xcolor}
\usepackage[normalem]{ulem}
\usepackage[nodayofweek,level]{datetime}

\usepackage{mathtools}
\usepackage{bm}
\usepackage{amssymb}

\usepackage[british]{babel}

\usepackage[square,numbers,comma,sort&compress]{natbib}
\title{\bf Imperfectly coordinated water molecules pave the way for homogeneous ice nucleation}

\usepackage[blocks]{authblk}
\author[1\dag]{Mingyi Chen}
\author[1\dag]{Lin Tan}
\author[2,3]{Han Wang}
\author[4]{Linfeng Zhang}
\author[1*]{Haiyang Niu}
\affil[1]{State Key Laboratory of Solidification Processing, International Center for Materials Discovery, School of Materials Science and Engineering, Northwestern Polytechnical University, Xi’an 710072, P.~R.~China}
\affil[2]{Laboratory of Computational Physics, Institute of Applied Physics and Computational Mathematics, Beijing 100094, P. R.~China}
\affil[3]{HEDPS, CAPT, College of Engineering, Peking University, Beijing 100871, P.~R.~China}
\affil[4]{DP Technology, Beijing 100080, P. R. China; AI for Science Institute, Beijing 100080, P.~R.~China}
\affil[*]{e-mail: haiyang.niu@nwpu.edu.cn}
\affil[ \dag]{These authors contributed equally}

\date{\formatdate{15}{8}{2022}}

\begin{document}

\maketitle

\begin{abstract}

Water freezing is ubiquitous on Earth, affecting many areas from biology to climate science and aviation technology\cite{bartels2012,bartels2013,brini2017water}. Probing the atomic structure in
the homogeneous ice nucleation process from 
 scratch is of great value but still experimentally unachievable\cite{sellberg2014,gallo2016water}. 
Theoretical simulations have
found that ice originates from the low-mobile region with increasing abundance and persistence of tetrahedrally coordinated water molecules\cite{matsumoto2002,tanaka2012,moore2011a,bullock2013,overduin2015,niu2019,fitzner2019}.
However, a detailed microscopic picture of how the disordered hydrogen-bond network rearranges itself into an ordered network is still unclear. 
In this work, we use a deep neural network (DNN) model to ``learn'' the interatomic potential energy from quantum mechanical data, thereby allowing for large-scale and long molecular dynamics (MD) simulations with \textit{ab initio} accuracy.
The nucleation mechanism and dynamics at atomic resolution, represented by a total of~$36\,\mu$s-long MD trajectories, are deeply affected by the structural and dynamical heterogeneity in supercooled water.
We find that imperfectly coordinated (IC) water molecules with high mobility 
pave the way for hydrogen-bond network rearrangement,  
leading to the growth or shrinkage of the ice nucleus.
The hydrogen-bond network formed by perfectly coordinated (PC) molecules stabilizes
the nucleus, thus preventing it from vanishing and growing.
Consequently, ice is born through competition and cooperation between IC and PC molecules.
We anticipate that our picture of the microscopic mechanism of ice nucleation will provide new insights into many properties of water and other relevant materials.
\end{abstract}

\newpage

Supercooled water with a disordered hydrogen-bond (H-bond) network separates
itself from crystalline ice by a large energy barrier\cite{matsumoto2002,moore2011a,niu2019}. 
It is still beyond the usual experimental techniques to detect the microscopic mechanism of how
the disordered water molecules transform to an ordered crystalline state\cite{gallo2016water}.
Instead, considerable computational efforts have been devoted to understanding the dynamics of structure transformation
during the nucleation process\cite{matsumoto2002,tanaka2012,moore2011a,bullock2013,overduin2015,niu2019,fitzner2019}. 
The first, which remains the only, homogeneous nucleation trajectory using
brute-force molecular dynamics (MD) with the empirical water model, i.e., TIP4P,
has shown that a fairly compact initial nucleus of ice forms on the occasion of a sufficient number of long-lived H-bonds\cite{matsumoto2002}. 
Simulations using both all-atom TIP4P/ice\cite{niu2019,fitzner2019} and coarse-grained mW\cite{moore2011a} water models have
clarified 
that ice originates from the increased abundance 
and persistence of the four-coordinated regions. 
Such a region has a high degree of tetrahedrality and low mobility\cite{tanaka2012,fitzner2019}, and ice was believed to be born in it\cite{fitzner2019}. However, many mysteries still linger regarding the ice nucleation process.
The main puzzle is how the dynamical and volatile disordered H-bond network rearranges, especially the mechanism of H-bond formation/cleavage, into an ordered state. 
In addition, as a critical experimental measurable property of water, the ice nucleation rate serves as a perfect quantity to probe the kinetics of water and build the link between experimental and computational studies\cite{haji2015}. 
However, the directly calculated nucleation rates using empirical water models, e.g., TIP4P/ice and mW, under homogeneous nucleation conditions are usually in poor agreement with the experimental values\cite{sosso2016}.
The discrepancy may be due to the simplification of the complex intra- and intermolecular interactions of water molecules in empirical water models that lead to inaccurate descriptions of the structure and dynamics\cite{kadaoluwa2021,haji2015}. Therefore, modelling the ice nucleation process at molecular level with sufficient accuracy is essential to reach a thorough microscopic mechanism of ice nucleation.

\begin{figure}
  \centering
  \includegraphics[width=0.95\columnwidth]{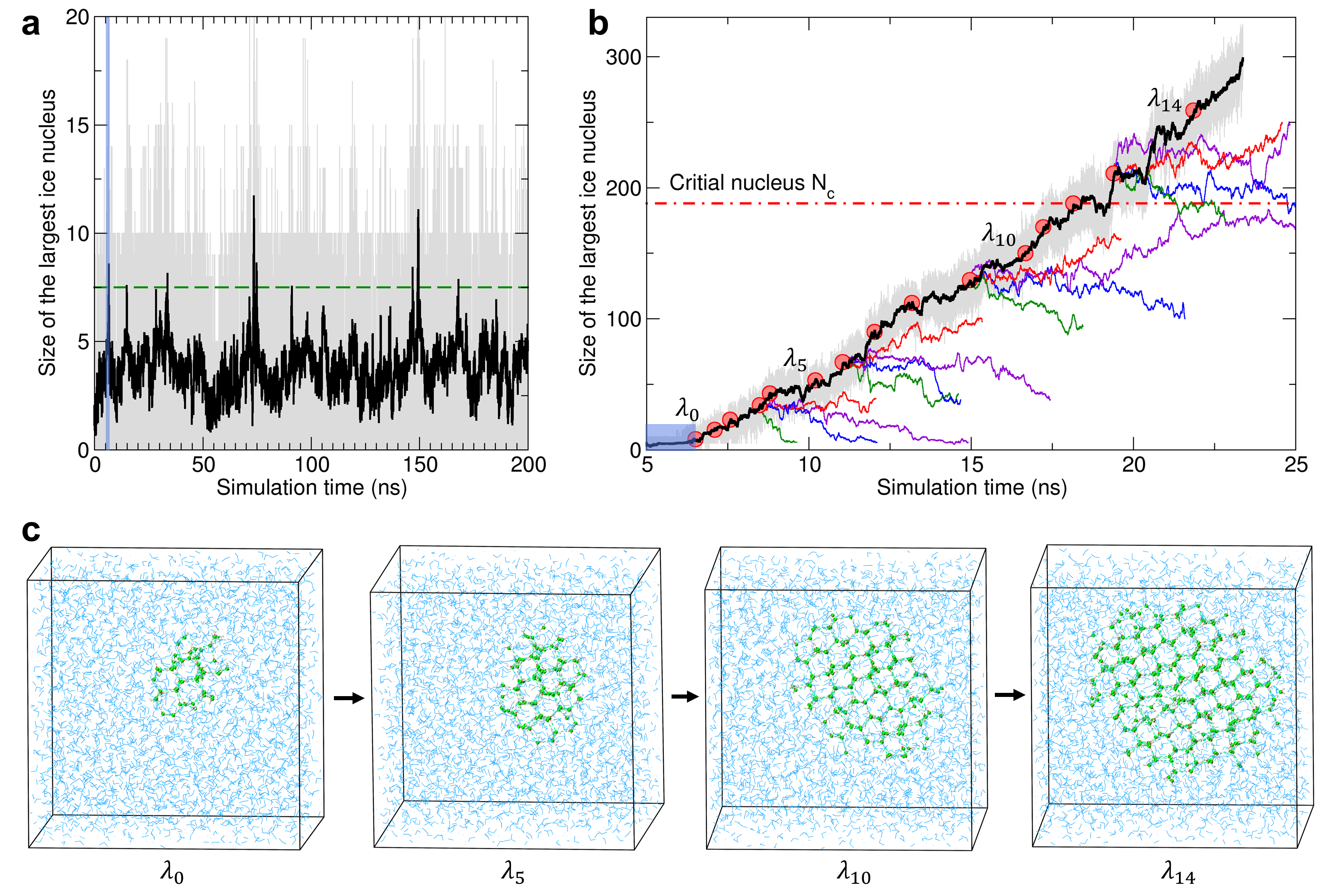}
  \caption{ \textbf{Atomic-scale simulation of the homogeneous nucleation process of ice.}
  \textbf{a,} Size of the largest ice nucleus versus simulation time of 
  the trajectory started with pure liquid at supercooling 46\,K. The value expresses the ice molecule number in the largest nucleus, identified 
  by common neighbour analysis\cite{maras2016} and the cluster search algorithm.
  The grey line is the instantaneous nucleus size, and the black
  line is the exponential moving average value of the nucleus size, 
  as shown in \textbf{b}.
  The green dashed line is the first milestone ($\lambda_0$) of the FFS process.
  \textbf{b,} One pseudo-brute-force ice nucleation trajectory obtained with the FFS approach.
  The beginning part of the trajectory refers to the one in \textbf{a}, highlighted by a blue shadow.
  Red dots represent the set of milestones from the first  ($\lambda_0$) to the fifteenth ($\lambda_{14}$).
  Some subtrajectories are shown in the plot.
  \textbf{c,} Snapshots of several ice nuclei over time. The corresponding points in \textbf{b} have been marked.
  }
  \label{F:cv}
\end{figure}

\textit{Ab initio} MD accurately computes the molecular interaction by solving the ground state electronic structure via the density functional theory (DFT) framework\cite{chen2017a}, however, due to the high computational cost, it is not capable of simulating ice nucleation from thousands of supercooled water molecules.
In this work, we overcome this challenge using a deep neural network (DNN) water model\cite{zhang2021}, which enables large-scale and long MD simulations with \textit{ab initio} accuracy by learning from DFT data calculated with the SCAN approximation of the exchange-correlation functional\cite{sun2015strongly,sun2016accurate}.
The choice of configurations used in the training set is crucial to maintain the high precision of the DNN model, 
and enhanced sampling-based MD simulations 
help to collect all the physically relevant configurations\cite{niu2020ab,bonati2018}. 
Here, we extend the training set of the DNN model by adding characteristic structures covering the phase space of the liquid, the crystalline, and especially, the nucleation regions in which the two phases coexist, obtained through metadynamics simulation. 
The melting temperature $T_{m}$ of the refined DNN water model is estimated to be 311\,K, which is higher than the experimental value but in good agreement with previous theoretical results using the same SCAN functional\cite{zhang2021,piaggi2021phase}.
Using the DNN water model, we performed MD simulations with a system of 2,880 water molecules 
at supercooling $\Delta T$ = 46\,K (based on the melting point of the DNN model) around the homogeneous nucleation conditions\cite{moore2011a}.

\section*{Homogeneous ice nucleation trajectories}

Fig.~\ref{F:cv} presents the evolution of the largest ice nucleus size throughout the obtained trajectories.
Within the 200\,ns trajectory shown in Fig.~\ref{F:cv}a, no nucleation event is observed. 
In fact, simulating one nucleation event is expected to take approximately 0.1\,s under the considered supercooling condition, which is far beyond the available temporal scale (typically $10^{-6}$\,s) of MD simulations.
However, several precritical 
ice nuclei with an instantaneous size of over 15 molecules form.
The key is how to nurture such nuclei to larger ones exceeding the critical size to obtain a successful nucleation event.
We choose the forward flux sampling (FFS)\cite{haji2015} method to overcome the time-scale limit of atomic simulations.
In the FFS approach, a collective variable (CV) defined as the largest ice nucleus size (molecule number) is monitored during the MD simulations. 
The nucleation procedure is divided into multiple stages by milestones that are defined by a list of monotonously increasing CV values $\{\lambda_0,\lambda_1,\dotsb,\lambda_n\}$.
At each milestone, dozens to hundreds of MD simulations starting from multiple initial configurations are launched, and they all have a chance to reach the next milestone. 
In this manner, continuous pseudo-brute-force nucleation trajectories can be obtained from 
a total of 36\,$\mu$s-long MD trajectories.

Fig.~\ref{F:cv}b plots a successful ice nucleation trajectory with the milestones highlighted by red dots. 
The initial configuration used at the first milestone has a nucleus of size 22 and comes from the trajectory in Fig.~\ref{F:cv}a.
Several snapshots of different ice nuclei labelled $\lambda_0$, $\lambda_5$, $\lambda_{10}$, and $\lambda_{14}$  are presented in Fig.~\ref{F:cv}c, 
and are marked in Fig.~\ref{F:cv}b at the corresponding time.
The ice nucleus is cubic (Ic) and hexagonal (Ih) stacking-disordered, which is consistent with previous experimental and theoretical observations\cite{murray2005,malkin2012,haji2015,lupi2017,pipolo2017,moore2011b,niu2019}.
Counterintuitively, the ice nucleus size shows a rugged and reversible fluctuating behaviour throughout the whole nucleation process. 
The ice nucleus size could change slightly in a relatively long term, i.e., over 1\,ns.
It can also grow or shrink dramatically by more than 20 molecules in a relatively short term, 
i.e., less than 100\,ps, as demonstrated by the subtrajectories launched at each milestone 
in the FFS simulation (Fig.~1b).

\begin{figure}
  \centering
  \includegraphics[width=0.95\columnwidth]{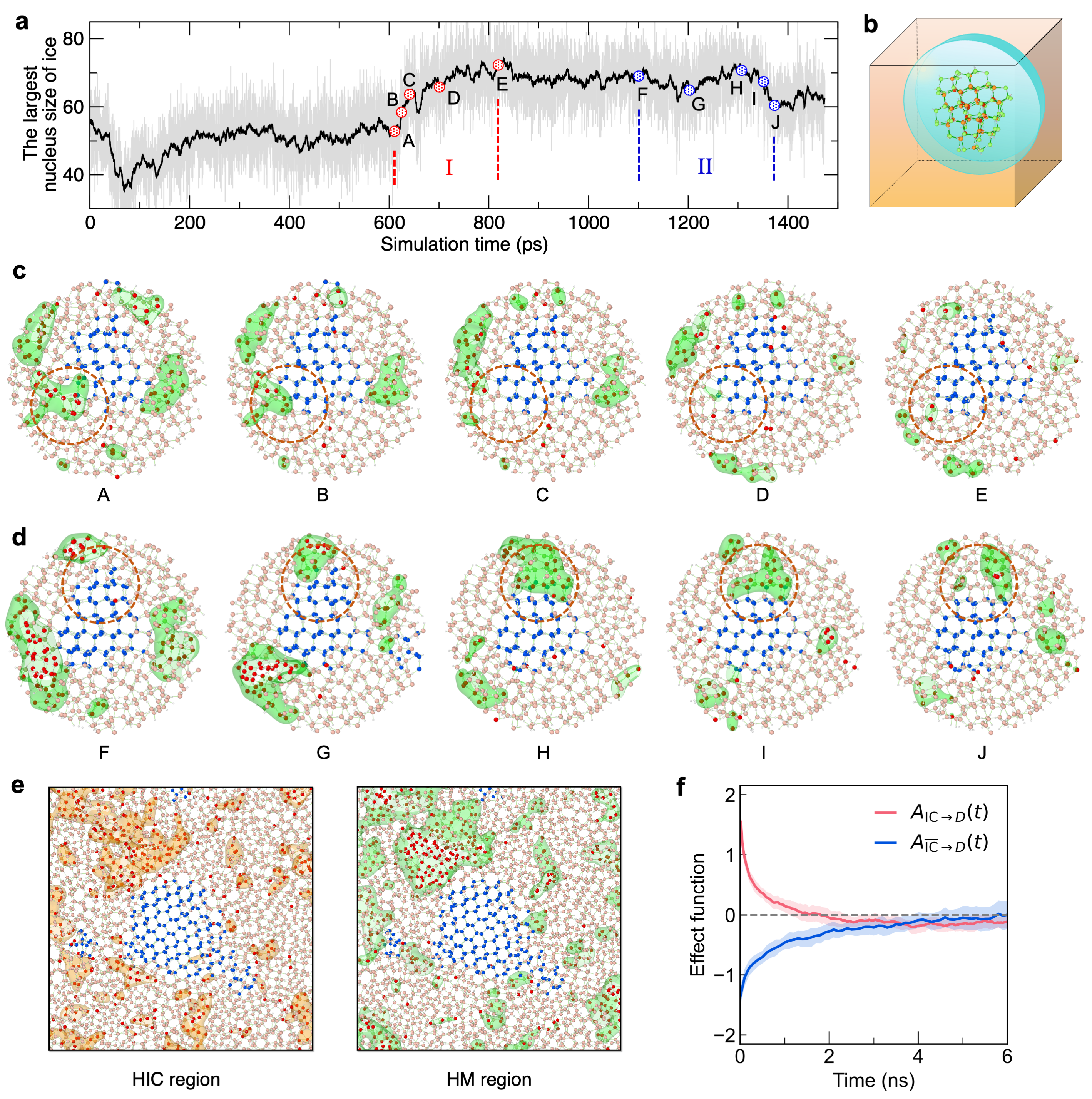}
  \caption{ \textbf{
 Dominant role of IC molecules in ice formation accompanied by structural and dynamical heterogeneity.}
  ~\textbf{a,} The largest nucleus size of ice versus the simulation time of one trajectory obtained in the FFS approach.
  Two representative processes, marked by I and II, refer to
  the growth and shrinkage of the nucleus, respectively. 
  \textbf{b,} The largest ice nucleus of the initial configuration  used in \textbf{a}. 
  A slab is cut out from the simulation box for detailed observation (the same for \textbf{c} and \textbf{d}).
  \textbf{c, d,} 
  Snapshots of the growth (\textbf{c}) and shrinkage (\textbf{d}) processes in \textbf{a}. 
   Blue molecules are ice detected by common neighbour analysis.
  Red molecules are HM molecules, and pink molecules have relatively low mobility. The regions wrapped by green surfaces are defined as HM regions.
  The instant of each snapshot has been marked in \textbf{a} with the corresponding capital letter.
  \textbf{e,} Correspondence between the HIC and HM regions. 
The regions wrapped by orange and green surfaces refer to the HIC and HM regions, respectively. The system illustrated here consists of 23040 water molecules.
  \textbf{f,} Effect function $A_{\mathrm{IC}\rightarrow D}(t)$ shows a strong correlation between
  imperfectly coordinated molecules near the nucleus surface and the change in nucleus.
  On the other hand, the negative effect shown by $A_{\overline{\mathrm{IC}}\rightarrow D}(t)$ indicates that PC molecules prevent the ice nucleus changes.
  }
  \label{F:mob}
\end{figure}

\section*{Role of imperfectly coordinated water molecules}

A visual inspection of the nucleation trajectories shows that the shape of the ice nucleus does not evolve in an isotropic manner; rather, some directions of the nucleus grow and shrink dramatically, while others remain unchanged in a relatively long term. 
Such dynamical anisotropic growth of ice nuclei might be
connected with dynamical heterogeneity,
a notable phenomenon of the coexistence of spatially distinguishable relatively mobile and immobile regions in supercooled water\cite{mazza2006,fitzner2019}.
The mobility of water molecules is contributed by
the translational and rotational degrees of freedom. 
Here, we use a measure to quantify the translational and rotational mobility (see Methods section for the definition). 
A threshold corresponding to the top 10\% of translational or rotational mobility in pure liquid water is set to distinguish the high-mobile (HM) molecules from others.
The regions mainly composed of HM molecules are defined as HM regions.
One representative trajectory in FFS involving both growth (process I) and shrinkage (process II) and the initial configuration are shown in Fig.~\ref{F:mob}a and \ref{F:mob}b, respectively. 
In process I (Fig.~\ref{F:mob}c), the HM region located at the bottom left of the nucleus in snapshot A transforms to ice in snapshot E, with the gradual depletion of HM molecules.
In process II (Fig.~\ref{F:mob}d), from snapshots F to H, an HM region near the top surface of the nucleus gradually expands to the ice nucleus and subsequently crushes the ordered H-bond network into a disordered state in snapshots I and J.
These results illustrate that the HM regions act on ice nucleation by prominent rearrangement of the H-bond network.
Significantly, the HM region does not necessarily introduce growth or melting of ice nucleus.
For example, in process II (Fig.~\ref{F:mob}d), the bottom left surface of the nucleus remains unchanged, although it is adjacent to an HM region, the H-bond network in this area has gone through dramatic reconstruction.
In contrast, other areas in processes I and II that are not part of or adjacent to the HM regions remain almost unchanged.

To understand the ice nucleation process from a structural perspective, we refer to the molecules forming four H-bonds, two accepting and two donating, with neighbouring molecules as perfectly coordinated (PC) molecules, and other molecules as imperfectly coordinated (IC) molecules. That is to say, an IC molecule can be viewed as carrying an IC status.
We measure the probability of molecules with IC status in a given time window and define the one with a high IC status probability as an HIC molecule.
Similar to the HM region, the HIC region can also be defined.
See the Methods section for the details of these definitions.
As shown in the left panel of Fig.~\ref{F:mob}e, the HIC region is spatially distinguishable from the remainder, indicating a structural heterogeneity behaviour, in accord with previous observations of the supercooled water structure\cite{sellberg2014,tanaka2012,gallo2016water}.
In the right panel of Fig.~\ref{F:mob}e, we highlight the HM regions and find an almost perfect agreement with the HIC regions.
To further investigate the correspondence of structural and dynamical heterogeneity, we calculate the cumulative distribution function of the translational and rotational mobility of molecules. A strong correlation  between the coordination status of water molecules and their mobility is shown, i.e., most IC molecules exhibit higher mobility than PC molecules, in agreement with previous work\cite{mazza2006,sciortino1991}. 
The time correlation function between HIC and HM molecules reveals a strong correlation.

From the discussion above, we can see that the role of the HIC regions in nucleus evolution is similar to that of the HM region, mainly due to the correspondence between structural and dynamical heterogeneity, resulting in the anisotropic growth or shrinkage of the ice nucleus. 
We further demonstrate the effect of HIC molecules on the nucleus by computing the effect function $A_{\textrm{IC}\rightarrow D}(t)$, defined by a net time-correlation between the appearance of IC molecules near the nucleus surface and the size change in the ice nucleus (see Methods for more details).
From Fig.~\ref{F:mob}f, a positive effect with a typical decay time of $\sim 1.5$\,ns is observed.
These results imply that the IC molecules, only 4.88\% of all molecules, act as engines to stimulate the HM region in supercooled water, 
drive the H-bond network rearrangement, and change the size of the ice nucleus.
On the other hand, the effect of a PC H-bond environment (no IC molecule in the neighbourhood of a PC molecule) on the nucleus size change is investigated by the effect function $A_{\overline{\textrm{IC}}\rightarrow D}(t)$.
A clear negative effect with a decay time of $\sim$4\,ns proves that the H-bond network formed by PC molecules prevents the nucleus from growing or vanishing.

It should be emphasized that such a role of IC molecules does not intend to pilot the structure \textit{a priori} to either the liquid or solid phase, which means that this is a ``dual role'' effect, and ice nucleation proceeds through competition and cooperation between IC and PC molecules.
When IC molecules drive growth or shrinkage in some parts of the nucleus, the H-bond network formed by PC molecules prohibits the nucleus from melting or growing in the directions where the IC molecules are scarce or absent.
This conclusion is different from the previous observations that ice was believed to be born in low mobile regions of supercooled water\cite{fitzner2019}, 
in which the abundance and persistence of the tetrahedrally coordinated water molecules act as the precursor structure of ice nucleation\cite{matsumoto2002,tanaka2012,moore2011a,bullock2013,overduin2015,niu2019}. 
The hidden side of the microscopic picture of ice nucleation is that PC molecules alone cannot nurse disordered water into crystalline ice, but IC molecules play the decisive role of paving
the pathway to ice formation. 
Our findings still hold if one traces back to the very beginning of nucleation that ice is born in high-mobile and low-mobile coexisting regions.

\begin{figure}
  \centering
  \includegraphics[width=0.9\columnwidth]{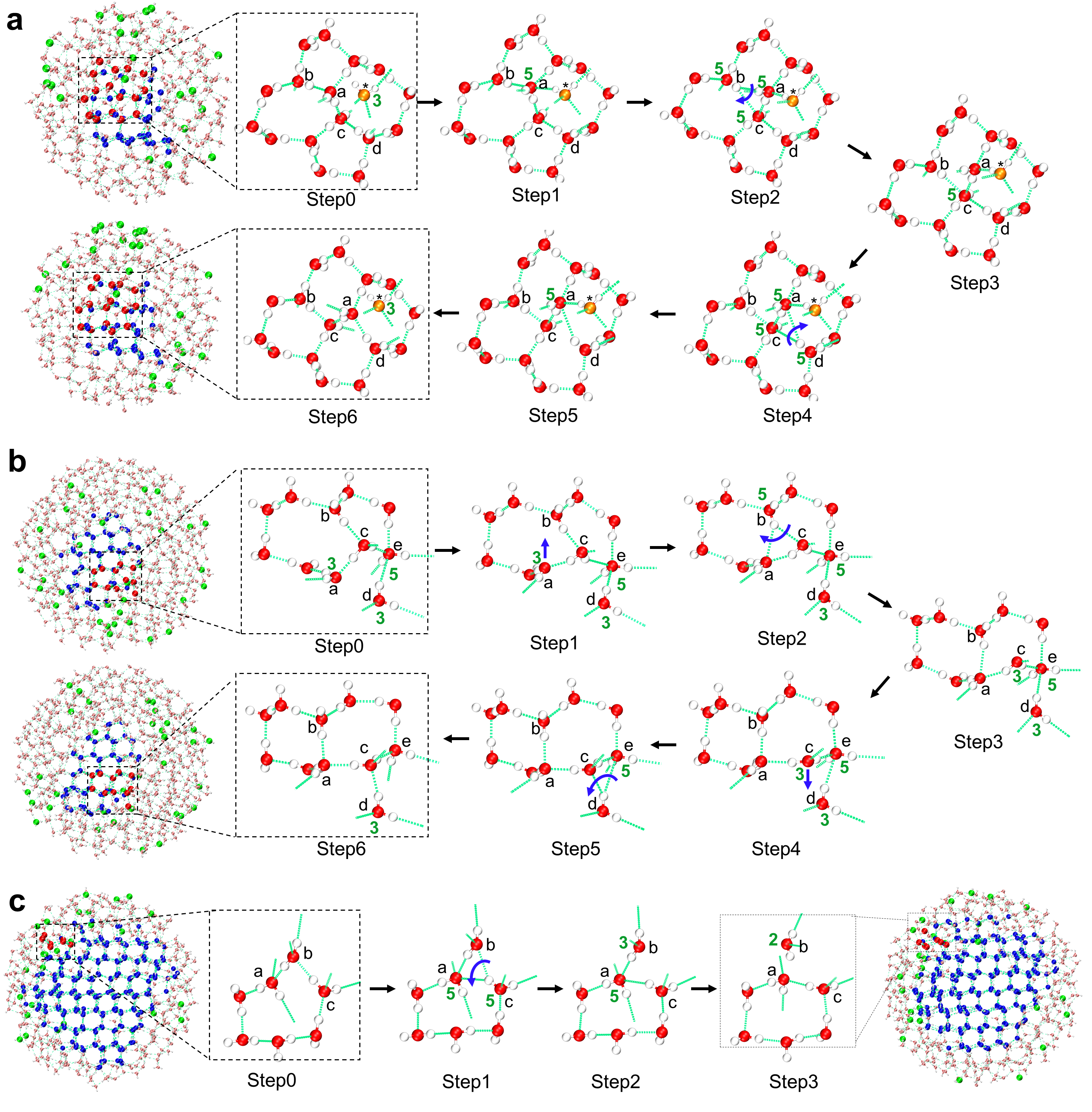}
  \caption{ \textbf{Microscopic mechanisms for H-bond network 
  rearrangement during ice formation.}
  Three representative ice growth events occur when one or several six-membered rings are constructed.
  \textbf{a,} Ice growth through IC status transfer. IC molecule W$^*$ acts like a ``catalyst'' that induces the transfer 
  of the imperfectly coordinated status among different molecules, facilitating the 
  formation of six-membered rings around it.
  \textbf{b,} Ice growth through IC status transfer and annihilation.
  The IC molecule W$^a$ first transfers its undercoordinate status to W$^c$, 
  and then this IC status annihilates with the overcoordinated  status of molecule W$^e$, 
  resulting in two PC molecules.
  \textbf{c,} Ice growth through IC status generation and transfer. Two PC molecules W$^a$ and W$^c$ lean close to each other and turn into IC molecules. 
  Then, W$^b$ drifts away from W$^a$ and W$^b$, recovering two PC molecules and 
  leaving one IC status on itself. 
  Specific water molecules are indexed with letters. The coordination number of IC molecules is marked with green numbers.
  Snapshots before and after the process are given, in which IC and PC molecules are coloured 
  in green and pink, respectively, ice-like molecules are coloured in blue, and
  hydrogen atoms are in light grey.  }
  \label{F:mech}
\end{figure}

\section*{Microscopic mechanism of H-bond network rearrangement}

To induce ice formation, water molecules must reorient markedly and break at least the H-bond involving the rotating hydrogen. In order to get a deeper insight into the role of IC molecules during ice formation, we, therefore, monitor the ice growth or shrinkage events, in which the six-membered rings of the ice nucleus that follow the Bernal-Fowler ice rule\cite{bernal1933} are formed or broken.

Fig.~\ref{F:mech} presents three typical events that show the microscopic mechanism of 
how disordered water transforms into ordered ice. 
In the following description, W with superscript denotes a specified water molecule. As shown in Fig.~\ref{F:mech}a, in Step 0, 
pairs of five- and seven-membered rings are composed of 16 PC molecules. 
One IC molecule, W$^*$, with three coordinations lies above 
these molecules. 
The W$^*$ first moves towards W$^a$ and forms a new H-bond with W$^a$ (Step 1). 
Now, the W$^*$ recovers the PC status, and W$^a$ is overcoordinated.
The H-bond from W$^b$ to W$^a$ oscillates until it bifurcates to form a new H-bond donated to W$^c$ (Step 2).
After that, W$^b$ reorients, and the H-bond between W$^a$ and W$^b$ breaks (Step 3).
The bifurcated H-bonds lower the activation energy of the H-bond network rearrangement from Steps 1 to 3 because the interaction energy of a bifurcated H-bond is roughly half of a linear bond\cite{giguere1984}.
As a consequence, W$^c$ becomes overcoordinated. 
In a similar procedure, the H-bond from W$^d$ to W$^c$ bifurcates to form a new H-bond to W$^a$ (Step 4), and then it breaks (Step 5).
Eventually, the five-coordinated molecule W$^a$ loses the H-bond with W$^*$, 
and W$^*$ returns to its initial three-coordinated status (Step 6).
Reviewing the whole process, one can find that it starts with the presence of the IC molecule W$^*$, which acts as a ``catalyst''. 
Such a catalysing molecule can be viewed as carrying one active defect, i.e., an IC status.
In this case, the IC status transfer follows a
W$^*$ $\rightarrow$ W$^a$ $\xrightarrow {\text{W$^b$}}$W$^c$ $\xrightarrow {\text{W$^d$}}$W$^a$ $\rightarrow$ W$^*$ path,
in which W$^b$ and W$^d$ act as bridges. 
At the very end, the two pairs of five- and seven-membered rings transform into four six-membered rings, 
resulting in a bigger ice nucleus.

Along with the nucleus size increasing, the liquid region shrinks in the simulation box. Therefore, the concentration of IC molecules in the liquid region should increase since the IC status never dissipates according to the IC status transfer mechanism discussed above.
However, our analysis shows that the concentration of IC molecules maintains dynamical equilibrium in the liquid region, indicating the annihilation and generation of IC status in supercooled water.
We find that ice formation events can occur through IC status annihilation or generation since such phenomena also lead to H-bond network rearrangement.
Fig.~\ref{F:mech}b presents a typical event involving the annihilation of IC statuses carried by two IC molecules. The IC status initially carried by W$^a$ transfers to W$^c$ (Steps 0 to 3). 
Then, the three-coordinated status of W$^c$ and five-coordinated status of W$^e$ annihilate, resulting in two PC molecules,
and finally, the initial pair of five- and seven-membered rings transforms into two six-membered rings (Steps 4 to 6). 
 
In contrast, Fig.~\ref{F:mech}c presents an ice formation event that is induced by IC status generation.
Here, molecule W$^a$, situated in the second neighbour shell of molecule W$^c$, fluctuates and occasionally moves near W$^c$, leading to the bifurcation of the H-bond donated from W$^c$ to W$^a$ and W$^b$ (Step 0 to 1). 
Subsequently, the H-bond from W$^c$ to W$^b$ breaks, and a pair of three- and five-coordinated molecules is generated (Step 2). 
Finally, the H-bond between W$^a$ and W$^b$ breaks, W$^a$ becomes PC, while W$^b$ changes to a two-coordinated status (Step 3).
This event leads a seven-membered ring to a six-membered ring.

By analyzing all the ice formation events throughout the whole ice nucleation trajectory in Fig.~1b,
we find that the H-bond network rearrangements are realized by three mechanisms, i.e., IC status transfer, annihilation, and generation, resulting in transformations between disordered and ordered states.
Among the three microscopic mechanisms, IC status transfer is the dominant player, approximately 90\% of the growth events rely on it.
In addition, events of ice nucleus shrinkage follow the same mechanisms in our observation. 
We find that the transfer process 
of IC status is different
from the molecular jump mechanism of water reorientation reported
in the supercooled water\cite{laage2006}; i.e., the rotating water molecule breaks an H-bond
with an overcoordinated first-shell neighbour and forms an H-bond with an undercoordinated second-shell neighbour. 
From our observations, in many cases, the reorientation and H-bond cleavage/formation can occur with the existence of a single IC molecule. It is worth noting that the ice nucleation microscopic mechanism presented here is also valid in the TIP4P/ice water model. 

\begin{figure}[p]
    \centering
    \includegraphics[width=1.00\columnwidth]{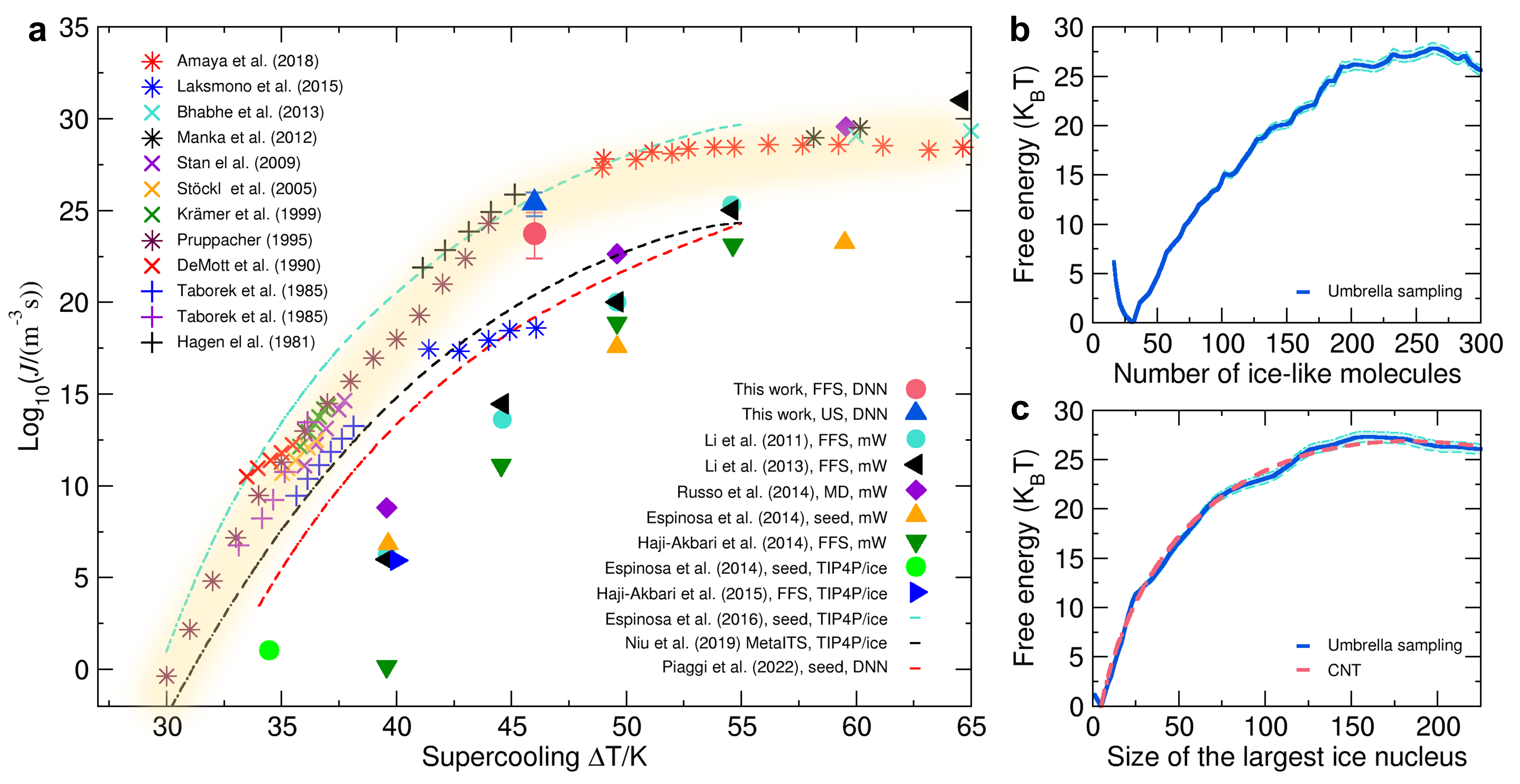}
    \caption{ \textbf{Homogeneous ice nucleation rate and free energy surface.}
    \textbf{a,} Ice nucleation rate dependent on the degree of supercooling, with 
    the ordinate logarithmically scaled based on 10. The data include the results 
    obtained in this work and some previous 
    theoretical\cite{li2011homogeneous,li2013ice,russo2014new,haji2014suppression,
    espinosa2014,haji2015,niu2019,piaggi2022}
    and experimental\cite{amaya2018ice,laksmono2015anomalous,bhabhe2013freezing,
    manka2012freezing,stan2009microfluidic,stockel2005rates,kramer1999homogeneous,
    pruppacher1995new,demott1990freezing,taborek1985nucleation} 
    work. 
    Experimental results are marked by crosses or asterisks with the legend located in 
    the upper left of the figure and theoretical results are solid polygons or 
    dashed lines with the legend in the lower right. 
    The yellow shadow region is drawn only to guide the eyes.
    \textbf{b,} Free energy surface constructed by umbrella sampling as a function of the
    number of ice-like molecules $N_i$, which is identified by the environment similarity metric\cite{piaggi2019a}. 
    The error bar is presented by the light-blue zone, 
    as shown in \textbf{b}.
    \textbf{c,} Free energy surface constructed by reweighting the umbrella sampling
    data to the axis of the size of the largest ice nucleus, the same collective 
    variable as in Fig.~\ref{F:cv}. 
    The red dashed line is obtained by fitting 
    the free energy curve using the function $f(n)=an^{2/3}+bn+c$ given 
    from the classical nucleation theory, where $a=3.200$ and $c=-7.475$ are
    obtained by approximating the blue curve,
   and $b$ is the chemical potential difference between the solid and liquid phases 
    equal to $-0.376$\,$k_BT$, which is calculated by thermal integration.
    }
  \label{F:fes}
\end{figure}

\section*{Ice nucleation rate}

Accurately estimating the ice nucleation rate under homogeneous conditions is another challenge for both experiments and computer simulations\cite{haji2015}. As shown in Fig.~4a, most experimental results\cite{amaya2018ice,laksmono2015anomalous,bhabhe2013freezing,
    manka2012freezing,stan2009microfluidic,stockel2005rates,kramer1999homogeneous,
    pruppacher1995new,demott1990freezing,taborek1985nucleation} fall into the yellow shadow region, while the theoretical results\cite{li2011homogeneous,li2013ice,russo2014new,haji2014suppression,
    espinosa2014,haji2015,niu2019,piaggi2022} are rather scattered.
Among all the theoretical techniques used to calculate the ice nucleation rate, the FFS method is known to be rigorous since it can calculate the rate directly. In this work, through the FFS method\cite{haji2015}, we obtain that the ice nucleation rate in logarithm is $\log_{10}J = 23.70^{+1.18}_{-1.32}$ ($J$ in $m^{-3}\cdotp s^{-1}$) and the critical nucleus size is $188$ at a supercooling of 46\,K. 
To validate the result obtained by the FFS method,
we further use the umbrella sampling method\cite{kastner2011} to calculate the free energy surface (FES, shown in Fig.~\ref{F:fes}b) along the nucleation path as
a function of the ice-like molecule number, identified by the environment 
similarity metric\cite{piaggi2019a}.
Then, we reweight this FES to the CV of the largest ice nucleus size and obtain Fig.~\ref{F:fes}c.
The red dashed line in Fig.~\ref{F:fes}c is the fitting curve using the classical nucleation theory (CNT). 
The homogeneous ice nucleation process conforms to the CNT reasonably well.
The critical size and the energy barrier obtained by the CNT fitting are $183^{+8}_{-8}$ and $26.85^{+1.45}_{-1.26}$\,$k_BT$, respectively.
Eventually, the logarithm of the nucleation rate is $\log_{10}J = 25.35^{+0.62}_{-0.68}$ ($J$ in $m^{-3}\cdotp s^{-1}$), which is consistent with the rate calculated by the FFS method, indicating the faithfulness of our estimation with both methods. 
The excellent agreement between our results and experiments substantiates the accuracy of the water model and the reliability of the ice nucleation microscopic mechanism proposed here.

\section*{Summary}

To summarize, we report MD simulation trajectories of 
ice nucleation under homogeneous conditions with the DNN water model that reaches \textit{ab initio} accuracy. 
We find that the IC molecules play a decisive role 
in ice nucleation in the sense that they pave the way between the disordered liquid state and the ordered solid state, and that the PC H-bond network 
preserves the existing ice nucleus from either growing or melting. 
Ice is born under competition and cooperation between the IC and PC molecules. 
Additionally, the nucleation process is well described by the CNT, and the nucleation rates estimated by this work are in excellent agreement with the experimental records. 
We expect that the relationship between IC molecules and nucleation in water provides new but important insight into the phase transition process and other complex behaviours of water. 
For other dynamic processes, such as heterogeneous ice nucleation, nucleation in water solution, and anti-freezing proteins, the effect of IC molecules should not be neglected. 

\bibliography{ms}

\section*{Methods}

\subsection*{Simulation programs and settings}

All molecular dynamics simulations are calculated on 
Lammps\cite{thompson2022}, integrated with 
DeePMD-kit\cite{wang2018} used to simulate with the deep neural network potential.
The simulation system is in an orthorhombic box with periodic boundary conditions 
in all three directions, which consists of 2880 water molecules unless otherwise specified.  
The time interval of the integral of motion, i.e., time step, is set to 0.5\,fs. 
Temperature and pressure are controlled by Nos\'e-Hoover style 
thermostat and barostat, set to 265\,K relaxed per 1\,ps and 1\,bar
relaxed per 0.05\,ps, respectively. Such temperature refers to a supercooling 46\,K of the DNN model.
Each particle starts with an initial velocity following the 
Maxwell-Boltzmann distribution and generated by a random seed.
In order to prevent the uniform rectilinear motion of the whole system, the linear 
momentum of each atom is rescaled by subtracting the momentum of centre of mass
every 1000 time steps. As for umbrella sampling and ice structure identification, 
Plumed 2\cite{tribello2014} is interfaced with Lammps so as to bias or analyse the trajectories. 
Crystal structure visualization is implemented by 
Ovito\cite{stukowski2010} and Vmd\cite{humphrey1996}.

\subsection*{Generation of the deep neural network (DNN) model}

The training of the DNN model is performed on the software package DeePMD-kit. 
The settings of training parameters are as follows: 
the size of the embedding net and fitting net are respectively (10,20,40) and 
(240,240,240). The cut-off radius is 6\,\AA. The smoothing parameter 
\textit{rcut\_smth} is set to 0.5\,\AA. In each training step, a minibatch, 
i.e., a small subset of the training dataset, is served to evaluate the gradient of the loss function. 
The hyper-parameters 
\textit{start\_pref\_e}, \textit{start\_pref\_f}, \textit{start\_pref\_v},
\textit{limit\_pref\_e}, \textit{limit\_pref\_f} and \textit{limit\_pref\_v} 
that control the weights of energy and force losses in the total loss function 
are set to 0.02, 1000, 0.05, 2.0, 1.0, and 0.5, respectively. 
The starting learning rate is 
0.002 and exponentially decays to $2\times10^{-8}$ at the end of the training. 
The total training steps are set to $1\times10^7$.

To build the DNN model applicable to water nucleation, 
our model has gone through an iterative update for better performance. 
The first generation of the DNN model is generated using the liquid configurations at standard temperature and pressure and all the experimentally known stable or metastable ice polymorphs for P $\lesssim$ 50\,GPa (including phases of Ih, Ic, II, III, IV, V, VI, VII, VIII, IX, XI, XII, XIII, XIV, XV). The total size of the dataset is 28097. All the above configurations are taken from Ref\cite{zhang2021}. 
However, such a dataset does not contain the 
solid-liquid coexistence configurations, which leads to insufficient accuracy of
the potential energy predictions during the nucleation process.
Therefore, 
we run metadynamics\cite{laio2002} simulations using the first generation of DNN potential, with a combination of the intensities of several structural factor peaks chosen as the collective variable\cite{niu2019}. 
In this way, 3734 new configurations consisting of liquid-Ic, liquid-Ih and 
liquid-Isd coexisting regions are added to the dataset, which is used to train the new generation of the DNN model that is used in this work.

\subsection*{Forward flux sampling}

Forward flux sampling (FFS)\cite{haji2015,allen2009} is a path sampling strategy in order to sample rare events dependent 
on a specified CV $\lambda$. Here, the whole nucleation process 
is divided into multiple stages. Each stage is marked by a certain value of CV.
Thereby a monotonous sequence $\{\lambda_0, \lambda_1, \dotsb, \lambda_n\}$ 
is established as ``milestones'' to assess the progress of the reaction.
The initial state (pure liquid) is separated by $\lambda \leq \lambda_l$, where $\lambda_l = 5$ is the smallest nucleus size that can be detected by the common neighbour analysis. Additionally, this value also corresponds to the lowest point on the free energy surface shown in Fig.~4b.
The main task of FFS is to estimate the probability 
for a trajectory starting from one milestone to reaching the next. The nucleation rate can be calculated by
\begin{equation}
  \label{eq:ffs}
  R = \frac{\overline{\Phi}_0}{\overline{h}_0}
  \prod_{i=0}^{n-1}P(\lambda_{i+1} | \lambda_{i}),
\end{equation}
in which $\overline{\Phi}_0/\overline{h}_0$
equals to the number of times of a long MD trajectory crossing 
$\lambda_0$ from $\lambda < \lambda_l$ divided by the length of simulation. 
$P(\lambda_{i+1} | \lambda_{i})$ is the transfer probability from milestone $\lambda_i$ to $\lambda_{i+1}$, 
estimated by performing a ``shooting test'' at $\lambda_i$,
which means to launch plenty of trajectories starting at $\lambda_i$ and 
calculate the probability of a trajectory reaching $\lambda_{i+1}$ 
but not falling into $\lambda_l$. 
The error of $P(\lambda_{i+1} | \lambda_{i})$ for each $i$ is evaluated by the standard deviation of 
transfer probabilities obtained from different initial configurations, and such error will propagate to the subsequent milestones along with Eq.~\eqref{eq:ffs}.

\subsection*{Umbrella sampling}

Umbrella sampling\cite{kastner2011} is one of the oldest but most versatile enhanced sampling methods 
based on free energy theory. In this method, the whole nucleation process is 
divided into multiple sections called ``windows''. The sampling is 
carried out on each window respectively. At last, merge the sampling data from each 
window and subsequently obtain the global free energy surface (FES). The sampling on each window is implemented by a long MD simulation that exerted an extra quadratic restraint.
The restraint added on sampling window $i$ can be written as 
\begin{equation}
    \label{eq:us}
    V_i = \frac{1}{2}K(s - s_i)^2,
\end{equation}
where $s$ is the CV, $s_i$ represents the centre of the $i$-th window, 
$K$ is a constant set to 1\,kJ/mol. The CV refers to the number of 
ice-like molecules, identified by the environment similarity metric. In total, 60 runs are performed to cover the CV space ranging from 20 to 315 with an interval of 5. For each run, a total of 16\,ns simulation is performed. 
We use the weighted histogram analysis method (WHAM)\cite{kumar1992a,zhang2019} to reweight the data and construct the FES. Eventually, a converged FES is obtained.

\subsection*{Ice structure identification and collective variables}

In this work, two methods are adopted to identify whether a water 
molecule is an ice molecule or not. 
For the process of FFS, the CV is chosen as the 
size of the largest nucleus (molecule number). To estimate it, 
an extended version of common neighbour analysis (CNA)\cite{maras2016} 
method is employed. 
This method takes the second nearest neighbours into consideration 
to discriminate the structures of cubic diamond and hexagonal diamond. 
In practice, only oxygen atoms are considered and in this 
way, each water molecule can be identified as Ih, Ic or others.
If a molecule is judged to be either Ih or Ic structure, its 4 nearest 
and itself are regarded as ice molecules. At the very end, a graph is established 
that each ice molecule links to its 4 nearest vertices, 
and the largest nucleus corresponds to the maximum connected component of the system. 
In order to reduce the computational costs, the calculation of the largest nucleus size
is performed every 200 time steps, i.e., 0.1\,ps.
Additionally, for convenient illustration, the nucleus size highlighted in all the snapshots includes the second shell of the central molecule.

However, this absolute ice nucleus size is non-differentiable and thereby 
cannot be used in biased molecular dynamical simulations. Hence, in the umbrella sampling, 
we adopt the environment similarity metric introduced by 
Piaggi et al.\cite{piaggi2019a}\ which is based on the 
methodology of smooth overlap of atomic position (SOAP)\cite{bartok2013}. This CV first 
compares the configuration $\omega$ presenting the
the environment of a central atom within a certain cut-off radius
respectively to each given configuration in the reference set 
$X=\left\{\chi_1, \chi_2, \dotsb, \chi_n\right\}$. 
In this work, only oxygen atoms are considered and $X$ consists of 
8 configurations that can completely describe the structure of 
Ih, Ic and Isd.
The similarity between $\omega$ and $\chi_k$ is calculated by
\begin{equation}
    \label{eq:soap}
    k_{\chi_k}(\omega) = \frac{1}{N_k}\sum_{i\in \omega}\sum_{j\in \chi_k}%
    \exp\left(-\frac{\lVert\bm{r}_i - \bm{r}_j\rVert^2}{4\sigma^2}\right),
\end{equation}
where $N_k$ is the atom number in configuration $\chi_k$, $\bm{r}_i$ and $\bm{r}_j$
are the position vector pertaining to the configuration $\omega$ and $\chi_k$ 
respectively with the origin located on the central atom. $\sigma$ is the width of 
Gaussian aimed to limit the deviation relative to the reference configuration.
Here, we set $\sigma$ to 0.055.
Next, $k_X(\omega) = \max\limits_{\chi_k\in X} k_{\chi_k}(\omega)$ is defined as the similarity 
metric compared to the reference set $X$. If $k_X(\omega)$ is larger than a certain value, 
we say the central atom is ``ice-like''. The CV used in umbrella sampling in this work
is the number of ice-like atoms in the whole system. In practice, this value is calculated
using a soft threshold by
\begin{equation}
    \label{eq:cv2}
    s = \sum_{i} \left(1 - w(k_X(\omega_i))\right),
\end{equation}
where $s$ denotes the calculated CV, $\omega_i$ is the environment of atom $i$.
The summation is applied to all oxygen atoms in the system.
$w(\cdot)$ is the switching function with the 
form of rational type
\begin{equation}
    \label{eq:sw}
    w(x) = \frac{1-(x/x_0)^{12}}{1-(x/x_0)^{24}}.
\end{equation}
This function will decay from 1 smoothly to zero as $x$ increases. 
$x_0$ is set to 0.55 in this work.

\subsection*{Data smoothing}

Grey lines in Fig.~\ref{F:cv} and Fig.~\ref{F:mob} show the instantaneous 
value of max nucleus size calculated by CNA, exhibiting violent 
fluctuation in a relatively short term 
during the whole simulation. 
To obtain a relatively stable largest nucleus size, we calculate the exponential moving 
average (EMA) by
\begin{equation}
    \label{eq:ema}
    S_t^{EMA}=
    \begin{cases}
        S_t & t=0; \\ 
        \alpha S_t+(1-\alpha)S_{t-1}^{EMA} & t > 0,
    \end{cases}
\end{equation}
where $S_t$ is the instantaneous value at $t$, $\alpha$ is a constant 
and set to $2/501$ in this work.
The EMA makes the smoothed data as close as possible to the original data 
but the fluctuations are greatly alleviated. We confirm this treatment to be reasonable 
since any changes in the largest nucleus size can be detected accurately by the smoothed 
data.

As for the analysis of the ice formation mechanism in Fig.~\ref{F:mech}, 
the trajectory is smoothed by computing the average of the coordinates 
over a time window centred at the current time. It is worth noting that the choice of the smoothing window size and the time interval between frames should not be too large to lose the concrete process of the H-bond reconstruction event. 
In this work, the trajectory is dumped every 10\,fs, and the smoothing window size is set to 20 frames. 

\subsection*{Measure of dynamical behaviours}

Dynamic behaviours of water molecules discussed in this work consist of two parts: 
translation and rotation.
For the translation behaviour of molecule $i$ at time $t$, we use 
$\langle\lVert\Delta \bm{r}_i (t) \rVert^2\rangle$ 
as an estimator. Here, the average $\langle\cdot\rangle$ is taken over a given time window $t$ to $t+T$. More concretely, it can be written in discrete form by
\begin{equation}
  \label{eq:trans}
  \langle \lVert \Delta \bm{r}_i (t) \rVert^2 \rangle = \frac{1}{T} 
  \sum_\tau^T \lVert \bm{r}_i(t + \tau) - \bm{r}_i(t) \rVert^2,
\end{equation}
where $\bm{r}_i(t)$ and $\bm{r}_i(t+\tau)$ is the position of oxygen atom 
of molecule $i$ at the present time $t$ and the time after $\tau$, respectively.

As for rotation, the definition of the estimator is similar to translation in 
Eq.~\eqref{eq:trans}. We start by the calculations of the rotation vectors using the method in Ref\cite{mazza2006} by Mazza et al.
To measure the three independent degrees of freedom of the
rotation, we set three mutually perpendicular axes on each molecule $i$ 
with the origin located at the oxygen atom. The unit vectors of these 
three axes are named $\bm{p}_i^1$, $\bm{p}_i^2$ and $\bm{p}_i^3$, respectively. 
$\bm{p}_i^1$ denotes the angular bisector of two O-H bonds. 
$\bm{p}_i^2$ is normal to the H-O-H plane. $\bm{p}_i^3$ directs to 
the cross product of $\bm{p}_i^1$ and $\bm{p}_i^2$.
Next, $\delta\bm{\varphi}_i^j(t)$ is defined to evaluate the rotation of 
molecule $i$ around axis $j$ between time $t$ and $t+\Delta t$, 
which is calculated by
\begin{equation}
  \delta\bm{\varphi}_i^j(t)=\cos^{-1}\left[\bm{p}_i^j(t)\cdot\bm{p}_i^j(t+\Delta t)
  \right]\frac{\bm{p}_i^j(t)\times\bm{p}_i^j(t+\Delta t)}%
  {\lVert\bm{p}_i^j(t)\times\bm{p}_i^j(t+\Delta t)\rVert}.
\end{equation}
This means the modulus of $\delta\bm{\varphi}_i^j(t)$ is equal to the size of 
rotation angle and the direction is normal to the rotation plane. 
Then $\bm{\varphi}_i^j(t)$ is calculated as
\begin{equation}
  \bm{\varphi}_i^j(t)=
  \sum_\tau^t\delta\bm{\varphi}_i^j(\tau).
\end{equation}
Then, $\langle\lVert\Delta\bm{\varphi}_i^j(t)\rVert^2\rangle$ is calculated likewise by taking average over the time window $t$ to $t+T$, i.e.,
\begin{equation}
  \label{eq:rot}
  \langle\lVert\Delta\bm{\varphi}_i^j(t)\rVert^2\rangle=\frac{1}{T}\sum_\tau^T
  \lVert\bm{\varphi}_i^j(t+\tau)-\bm{\varphi}_i^j(t)\rVert^2.
\end{equation}
We use the sum of 
$\langle\lVert\Delta\bm{\varphi}_i^1(t)\rVert^2\rangle$, 
$\langle\lVert\Delta\bm{\varphi}_i^2(t)\rVert^2\rangle$
and $\langle\lVert\Delta\bm{\varphi}_i^3(t)\rVert^2\rangle$ to estimate the
rotation behaviour of molecule $i$.

The length of time $T$ 
in Eq.~\eqref{eq:trans} and \eqref{eq:rot} should be set manually.
In this work, $T$ is set to 100\,ps. This choice is comparable with 
the length of the
half-life decay time of $A_{\text{IC}\rightarrow D}(t)$ shown in Fig.~\ref{F:mob}f in the main text.

\subsection*{High mobile regions}

The definition of the high-mobile (HM) region is based on HM molecules.
To determine whether a molecule is HM or not, we first simulate pure liquid water and calculate the distribution of the translational and
rotational mobility over all molecules of all frames. The threshold of high mobility
is corresponding to the highest 10\% mobile molecules. 
At supercooling 46\,K, the values for translation and rotation are 
7.7\,\AA$^2$/ps and 12.8\,rad$^2$/ps, respectively. We define that a molecule as HM if it has either 
of two types of mobility above the respective threshold. Then, smearing normalized 
three-dimensional Gaussian density function on each HM molecule, any position in the simulation box will take 
a value by the summation of all added functions. The region wrapped by an isosurface is defined as the HM region. 
In this work, the width of Gaussian is set to 1.73\,\AA, and the isovalue is set to 1.

\subsection*{IC status probability}

To better explain the relationship between the IC status and mobility, we introduce the measure of the IC status probability.
For each water molecule, its value is calculated as the frequency of the molecule being in IC status in a time window centred at the current time.
The size of the time window is set to 100\,ps, the same as the definition of the measure of mobility.
To make this measure comparable with HM regions, we first calculate the fraction of HM molecules in liquid water at supercooling 46\,K. 
The threshold for high IC status probability (HIC) is set so that the fraction of HIC molecules is the same as the fraction of HM molecules. We obtain the threshold value for HIC is 9.79\%.
The region of the HIC in Fig.~\ref{F:mob} is plotted through the same method as the HM region but the Gaussian is smearing on the HIC molecules.

\subsection*{Effect function}

To investigate the effect of the imperfectly coordinated (IC) water molecules on the size change of the ice nucleus, we define the effect function $A_{\mathrm{IC}\rightarrow D}(t)$ as the difference between the following two time correlation functions: 
\begin{align}\label{eq:effect-func}
    A_{\textrm{IC}\rightarrow D}(t) =
    C_{(N_{\textrm{IC}}\cap S), D}(t)
    -
    C_{S, D}(t),
\end{align}
where the time correlation functions are defined by
\begin{align}\label{eq:corr-0}
    C_{N_{\textrm{IC}}\cap S, D}(t) & = 
    \frac{\big\langle N_{\textrm{IC}}^i(T) \cdot S^i(T)\cdot D^i(T+t) \big\rangle_{i,T}}
    {{\langle N^i_{\textrm{IC}}(T) S^i(T)\rangle_{i,T}\langle D^i(T)\rangle_{i,T}}},\\\label{eq:corr-1}
    C_{S, D}(t) & = 
    \frac{\big\langle S^i(T)\cdot D^i(T+t) \big\rangle_{i,T}}
    {\langle S^i(T)\rangle_{i,T}\langle D^i(T)\rangle_{i,T}}.
\end{align}
In Eqs.~\eqref{eq:corr-0} and \eqref{eq:corr-1}, $N_{\textrm{IC}}^i(T)$ denotes the number of IC molecules in the neighbourhood $\mathcal N(i)$ of molecule $i$;
$S$ and $D$ are indicator functions (definition introduced later) telling if molecule $i$ is near the nucleus surface and if its status (ice or water) changes, respectively;
the average $\langle \cdot \rangle_{i,T}$ is taken over the molecular index $i$ and the simulation time $T$. 
The $C_{S, D}$ denotes the time-correlation function between indicators $S$ and $D$, while $C_{(N_{\textrm{IC}}\cap S), D}$ is the time-correlation function between an indicator function telling if a near-nucleus-surface water molecule
has IC molecules in its neighbourhood (denoted by $N_{\textrm{IC}}\cap S $) and the indicator function $D$.

To properly define the indicators, we introduce a status function $I^i(T)$ that takes the value of 1 if the molecule $i$ is belonging to an ice nucleus at time $T$, otherwise takes a value 0. 
The surface indicator $S^i(T)$ is defined by the standard deviation of the status function in the neighbourhood of molecule $i$, i.e.,
\begin{align}
    S^i(T) = \textrm{std}\big( I^j(T) \big)_{j\in \mathcal N(i)}.
\end{align}
It is clear that if the molecule is inside the water or ice region, all the status function $I^j$ respectively takes value 0 or 1, so the standard deviation vanishes. 
On the other hand, if the molecule is near the surface, then there must be some ice and water molecules co-existing in the neighbourhood, which leads to a non-zero surface indicator.
The time-dependent functions $N_{\textrm{IC}}^i(T)$ and $S^i(T)$ are smoothed along the trajectories with a window size of 40\,ps.

The change-of-status indicator is defined as the difference between the status function of two consecutive times along the trajectory separated by a time resolution $\tau$, i.e.
\begin{align}
    D^i(T) = \vert I^i(T+\tau) - I^i(T) \vert
\end{align}
The indicator function takes the absolute value of the difference in $I^i$ to take into consideration both the water-to-ice and ice-to-water changes. 
The indicator function $I^i(T)$ is smoothed along the trajectories with a window size of 40~ps.

Similar to the effect function $A_{\textrm{IC}\rightarrow D}(t)$, we define
$A_{\overline{\textrm{IC}}\rightarrow D}(t)$ quantifying
the effect of a PC environment (no IC molecule presents in the neighborhood of a molecule) and the change of the status of a molecule.
\begin{align}\label{eq:effect-func-1}
    A_{\overline{\textrm{IC}}\rightarrow D}(t) =
    C_{(\overline N_{\textrm{IC}}\cap S), D}(t)
    -
    C_{S, D}(t),
\end{align}
where $\overline N^i_{\textrm{IC}}$ is defined by
\begin{align}
    \overline N^i_{\textrm{IC}}(T) = \left\{
    \begin{aligned}
    1 \quad& \textrm{if}\  N^i_{\textrm{IC}}(T)=0;\\
    0 \quad& \textrm{otherwise}.
    \end{aligned}
    \right.
\end{align}

\subsection*{Calculation of nucleation rate based on classical nucleation theory}

Using the FFS method one can obtain nucleation rate directly, but the umbrella sampling can not.
Through the umbrella sampling the nucleation barrier $\Delta G^*$ is obtained. 
The classical nucleation theory (CNT)\cite{pruppacher2010} 
predicts the nucleation rate $J$ as
\begin{equation}
    \label{eq:cnt}
    J = \rho_l Z f^+ \exp\left(-\frac{\Delta G^*}{k_B T}\right),
\end{equation}
where $\rho_l$ is the number density of liquid, $Z$ is the Zeldovich factor, 
$k_B$ is the Boltzmann constant, $T$ is the temperature, $f^+$ is the attachment rate describing 
the diffusion behaviour near the critical 
nucleus size. The Zeldovich factor can be calculated by
\begin{equation}
    \label{eq:zeldovich}
    Z = \left[\frac{\Delta G^*}{3\pi k_B T N_c^2}\right]^{1/2},
\end{equation}
where $N_c$ is the number of molecules when the nucleus is critical. 
This equation of the Zeldovich factor is applicable to various 
geometry shape of the nucleus, including ellipsoid. 
The attachment rate is calculated as the diffusion coefficient of
the molecule number of the critical nucleus, written as 
\begin{equation}
    \label{eq:f}
    f^+ = \frac{\langle(N(t) - N(0))^2\rangle}{2t},
\end{equation}
where $\langle\cdot\rangle$ denotes the ensemble average and 
$N(t)$ is the molecule number of the nucleus at time $t$.
Here, the attachment rate $f^+$ of water at supercooling 46\,K is evaluated to be $3.4713\times 10^{10}$\,s$^{-1}$.

\section*{Data availability}
The main data supporting the findings of this study are available within the paper. Additional information and data are available from the corresponding author upon reasonable request.

\section*{Code availability}
LAMMPS, Plumed, and DeePMD-kit are free and open source codes available at https://lammps.sandia.gov, https://www.plumed.org, and http://www.deepmd.org, respectively. Additional codes are available from the corresponding author upon reasonable request.

\section*{Acknowledgements}

The authors would like to thank Prof.~Michele Parrinello, Prof.~Roberto Car, Prof.~Artem R.~Oganov, Prof.~Lars Stixrude, Dr.~Jie Deng, and Xiangyang Liu for many valuable discussions. H.N.~was supported by The National Natural Science Foundation of China (grant No.\,22003050) and the Research Fund of the State Key Laboratory of Solidification Processing (NPU), China (grant No.\,2020-QZ-03).
H.W.\ was supported by the National Science Foundation of China (grant No.11871110 and 12122103).
The calculations were supported by National Supercomputing Center in Zhengzhou and Tianjin, China, Bohrium Cloud Platform at DP technology, and the International Center for Materials Discovery (ICMD) cluster of NPU.

\section*{Author contributions}

H.N.\ conceived the idea and designed the research. H.W.\ conceived the idea to sample the pseudo-brute-force ice nucleation trajectories. M.C., L.T. and H.N.\ carried out the FFS simulations and analyzed the data. M.C.\ and H.N.\ carried out the umbrella sampling simulations. L.T., H.N.\ trained the water DNN model with the help of H.W.\ and L.Z.. H.W.\ performed the effect function analysis. M.C.\ and H.W.\ wrote the codes for data analysis. The text was initially composed by M.C., H.N, and H.W.\ and all the authors contributed to the discussion of the research and the final version of the manuscript.

\end{document}